\documentclass[twocolumn,superscriptaddress,aps]{revtex4}
\usepackage[greek,english]{babel}
\usepackage{graphicx}
\usepackage{amsmath}
\usepackage{mathrsfs}
\usepackage{amsfonts}
\usepackage{amssymb}

\begin{document}

\title{Seismic waves damping with arrays of inertial resonators}

\author{Younes Achaoui*}
\affiliation{Aix-Marseille Universit\'e, CNRS, Centrale Marseille, Institut Fresnel UMR 7249, 13013 Marseille, France}
\author{Bogdan Ungureanu*}
\affiliation{Aix-Marseille Universit\'e, CNRS, Centrale Marseille, Institut Fresnel UMR 7249, 13013 Marseille, France}
\affiliation{Faculty of Civil Engineering and Building Services
Technical University "Gheorghe Asachi" of Iasi,
43, Dimitrie Mangeron Blvd., Iasi 700050, Romania,}
\author{Stefan Enoch}
\affiliation{Aix-Marseille Universit\'e, CNRS, Centrale Marseille, Institut Fresnel UMR 7249, 13013 Marseille, France}
\author{St\'ephane Br\^ul\'e}
\affiliation{Dynamic Soil Laboratory, M\'enard, 91620 Nozay, France,}
\author{S\'ebastien Guenneau}
\affiliation{Aix-Marseille Universit\'e, CNRS, Centrale Marseille, Institut Fresnel UMR 7249, 13013 Marseille, France}

\date{\today}

\begin{abstract}
We investigate the elastic stop band properties of a theoretical cubic array of iron spheres connected to a bulk of concrete via iron or rubber ligaments. Each sphere can move freely within a surrounding air cavity, but ligaments couple it to the bulk and further facilitate bending and rotational motions. Associated low frequency local resonances are well predicted by an asymptotic formula. We find complete stop bands (for all wave-polarizations) in the frequency range $[16,21]$ Hertz (resp. $[6,11]$ Hertz) for $7.4$-meter (resp. $0.74$-meter) diameter iron spheres with a $10$-meter (resp. $1$-meter) center-to-center spacing, when they are connected to concrete via steel (resp. rubber) ligaments. The scattering problem shows that only bending modes are responsible for damping and that rotational modes are totally overwritten by bending modes. Regarding seismic applications, we further consider soil as a bulk medium, in which case the relative bandwidth of the low frequency stop band can be enlarged through ligaments of different sizes that allow for well separated bending and rotational modes. We finally achieve some damping of elastodynamic waves from $8$ to $49$ Hertz (relative stop band of $143$ percent) for iron spheres $0.74$-meter in diameter that are connected to soil with six rubber ligaments of optimized shapes. These results represent a preliminary step in the design of seismic shields placed around, or underneath, the foundations of large civil infrastructures. 
\end{abstract}

\maketitle

\section{Introduction}
Inserting stone blocks between a structure and its foundation is one of the first attempts to reduce vibrations and this can be traced back to the sixth century BC. In 1950,
the first specialized damping devices for earthquakes were developed and civil engineers started to implement different mechanisms to mitigate seismic damage to buildings, but with relatively weak and non-earthquake resistant structures \cite{Reitherman}. Nowadays civil engineers protect buildings in particular with metal springs, ball bearings and rubber pads, all of them designed to mitigate the energy from seismic waves \cite{Reitherman}. In earthquake engineering, the structure could be designed with additional passive energy dissipation systems integrated in the frame structure (dampers, rubbers) or semi-active dampers \cite{symans}. However the objective of this article is to illustrate that buried structures in the soil (foundations, buried levels of the buildings, etc.), could directly act, in the near future, on the seismic signal itself coming at the base of the buildings. In that sense, we preliminary demonstrate with numerical models, that we achieve filtering of frequencies by means of a buried device made of resonating iron spheres. This capability of the device to filter frequencies offers wide range of applications in soil-structure interaction (ISS), particularly if the foundations are designed both for their bearing capacity and as local resonant elements.

\maketitle

\section{Concept of seismic protection}
In general, seismic waves are vibrations that travel through the Earth carrying energy released during an earthquake. Seismic waves transport energy from the focus and when their longitudinal (P) and transverse (S) components meet the surface, their energy is partly converted into waves that propagate along the Earth surface. Such Rayleigh waves, which have one compressional and one shear components, are characterized by wavelengths ranging from meters to decameters what corresponds to frequencies from a few tens of Hertz to less than $1$ Hertz, where the velocity of the wave is decreasing, since the higher frequency components are more effectively attenuated during wave propagation. The fact that these surface waves have a lower speed, hence typical wavelengths much smaller than underground (bulk) waves, makes them the most destructive seismic waves for civil infrastructures built on sedimentary soils, due to the well-known phenomenon of resonance disaster \cite{brule1,brule1b}.
Indeed, when seismic waves propagate through soft superficial alluvial layers or scatter on strong topographic irregularities, refraction or scattering phenomena may strongly increase the amplitude of ground motion. At the scale of an alluvial basin, seismic effects involve various phenomena, such as wave trapping, resonance of whole basin, propagation in heterogeneous media, and the generation of surface waves at the basin edge \cite{brule1,brule1b}. Due to the surface wave velocity in superficial and under-consolidated recent material (less than $100$ m.s$^{-1}$ to $300$ m.s$^{-1}$), wavelengths of surface waves induced by natural seismic sources or construction work activities are shorter than those of earthquake generated direct P (primary, i.e. longitudinal compressional) and S (secondary, i.e. transverse shear) waves (considering the $0.1$ to $50$ Hertz frequency range), from a few meters to a few hundreds of meters. These are of similar length to that of buildings, therefore leading to potential building resonance phenomena in the case of earthquakes \cite{brule1,brule1b}. 

Most common technical solutions for structural protection are dampers, wave scatterers, resonators, rubber isolation etc. \cite{symans}. Here, we propose a radically different concept based on low frequency stop bands associated with meter-scale inertial resonators that might help with preventive measures of earthquake risk management in civil engineering and building structures.
Indeed, recent scientific advances in total or partial reflection band gaps (due to periodicity) and invisibility (artificial anisotropy via transformation physics) for elastic waves in plates, as well as wave magnitude damping (using Helmholtz's resonators), have opened new avenues in seismic metamaterials. A way to counteract the devastating effect of seismic waves is to design the shape of inclusions in the soil in order to tune stop bands in accordance with building's fundamental period and its harmonics. Although it is illuminating to draw analogies with control of elastic waves in plates, we have to take into consideration that seismic protection is much more complex than the dispersion/molding of the Lamb waves in thin, heterogeneous plates whose elastic properties are more straightforward to model and characterize. This, in particular, is due to the complex properties of the soil (viscoelastic, anisotropic and irregular propagation medium).

Mechanical resonance is the tendency of a building to respond with a larger amplitude when the frequency of its oscillations matches the building's natural frequency of vibration (resonant frequency). It may cause violent swaying and even catastrophic failure - a phenomenon known as resonance disaster \cite{Reitherman}. These resonant frequencies of the buildings can be calculated and are indeed taken into account in the design of their structures. The analysis could be carried out for free vibrations with an idealized single degree of freedom oscillator (S.D.O.F.O with oscillators). The global matrix expression describing the displacement of a multi-oscillator system, neglecting the damping matrix for a pre-design study, leads to the equation $\mid K-\omega_n^2 M \mid=0$ with $K$ the stiffness matrix in a linear case of the lateral force, $M$, the mass matrix and $\omega_n=2\pi/T_n$ the eigenfrequencies associated with the building's periods $T_n$ \cite{brule1,brule1a,gmur,chopra}. Using S.D.O.F.O typical eigenfrequencies of a five storey concrete building are found to be in the range $1$ to $100$ Hertz.

The first in situ experiment of a deflection shield for seismic waves led by the M\'enard company in collaboration with the Fresnel Institute near the Alpine city of Grenoble in 2012 has unveiled some similarities between acoustic waves in plates and surface seismic waves \cite{brule2,brule3} and this prompted researchers to envision the large-scale analogs of acoustic metamaterials \cite{sheng,djafari,mechanical1,mechanical2,li,movchan,fang,milton,sar,theocharis,hsu,craster,mechanical3,fleury}, with a coined name seismic metamaterials \cite{brule2}. Similarly to multiple scattering and locally resonant effects experimentally observed in microsonic and hypersonic phononic crystals around $1$ GHz \cite{achaoui2,achaoui3,achaoui1}, experimental results in \cite{brule1b,brule2,andrea1} have validated the possibility to prohibit frequencies around $50$ Hertz i.e. at the upper frequency range of interest as suggested by the S.D.O.F.O. model and further paved the way to explore locally resonant subwavelength structures that might improve such Bragg-based seismic shields.

The purpose of the present letter is to highlight the mechanisms offered by a new class of seismic metamaterials based on subwavelength inclusions that may enable a drastic attenuation of earthquakes. These can be achieved by engineering the frequency stop bands dedicated to seismic waves by converting the latter to evanescent waves in order to protect a sensitive area (schematics shown in figure 1).
We numerically demonstrate that seismic waves can be molded by changing the field propagation properties, by placing a cubic array of resonators in the seismic wave path.

\section{Low frequency stop bands with inertial resonators}

\begin{figure}[b]
\centering
\includegraphics[clip,angle=0,width=85mm]{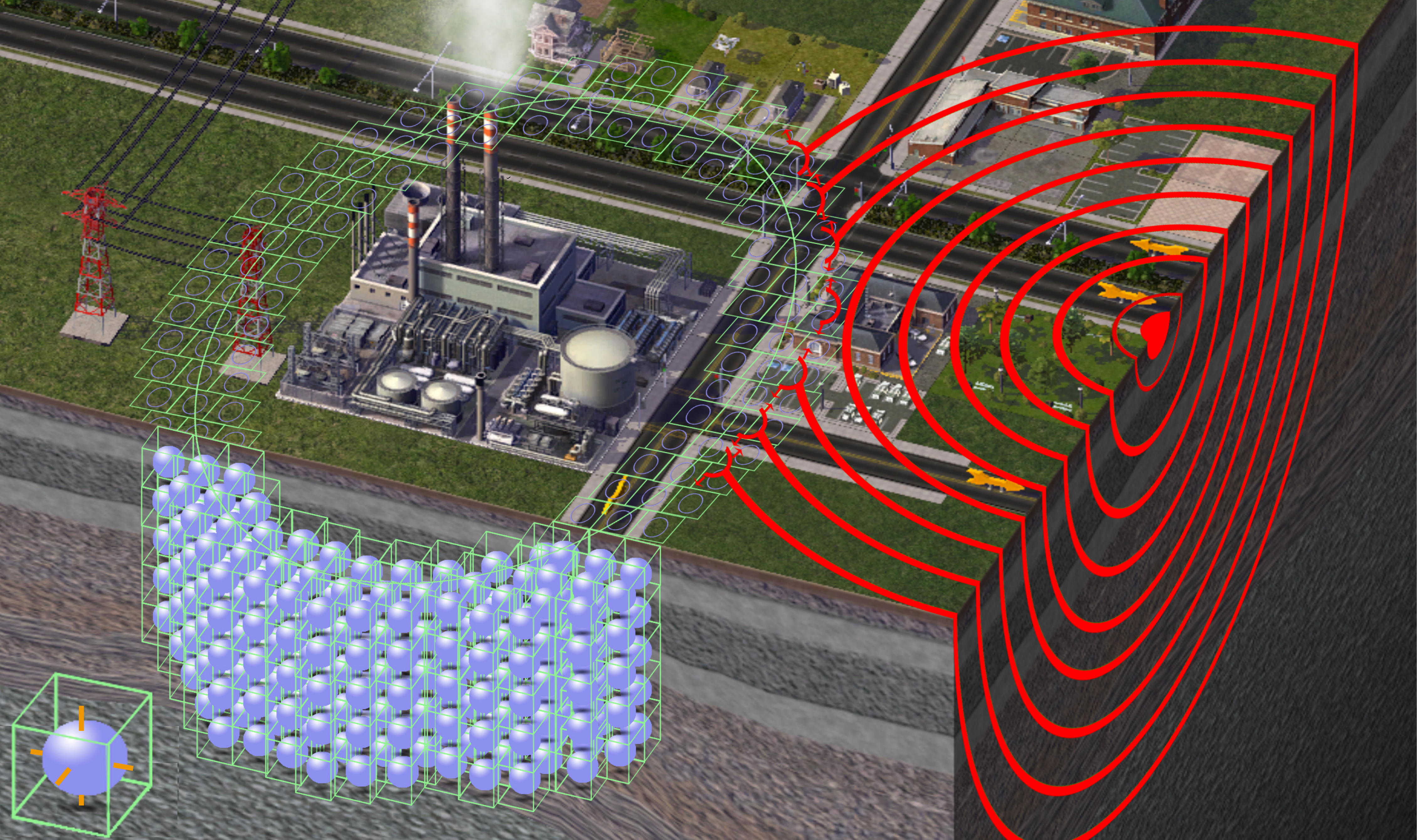}
\caption{Schematic diagram of a seismic wave shield consisting of inertial resonators placed around the foundations of a large civil infrastructure. Insert shows a periodic cell with an iron sphere connected to a bulk of concrete or soil through six iron or rubber ligaments.
}
\end{figure}

\begin{figure*}[t]
\centering
\includegraphics[clip,angle=0,width=170mm]{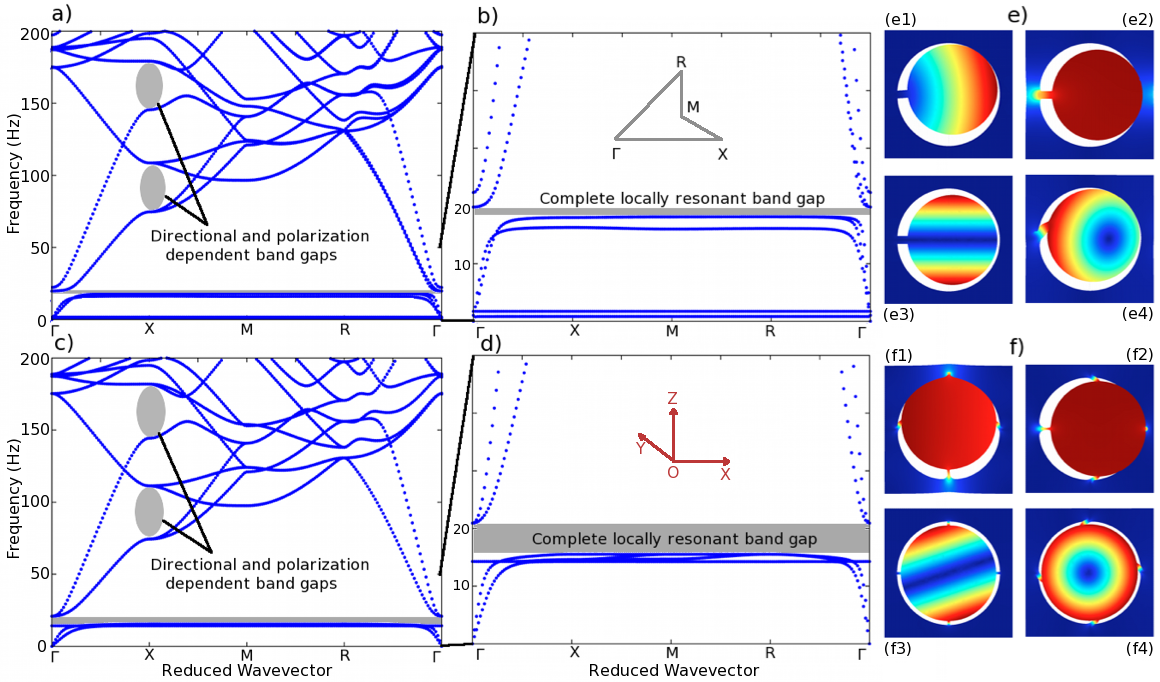}
\caption{Band diagrams for elastic wave propagation within a periodic structure with a cubic elementary cell 10 meters in sidelength made of an iron sphere of diameter $7$ m (resp. $7.4$ m) connected to concrete (host medium) by one iron ligament of length $0.8$ m and diameter $0.6$ m (a, b) and resp. six identical iron ligaments of length $0.3$ m and diameter $0.2$ m (c, d). (b) and (d) are close-up views of low frequency stop bands around $19$ Hertz. (e) and (f) are representative bending (e1,f1), compressional (e2,f2) and rotational (e3,e4,f3,f4) eigenmodes shown as inserts. All the inserts are displayed in the (Oxz) plane (sagittal cut) at the X point of the reduced Brillouin zone,
$\Gamma XMR$ with $\Gamma=(0,0,0)$, $X=(\pi/a,0,0)$, $M=(\pi/a,\pi/a,0)$, $R=(\pi/a,\pi/a,\pi/a)$, $a=10$ m being the array pitch. 
}
\end{figure*}

\begin{figure*}[t]
\centering
\includegraphics[clip,angle=0,width=170mm]{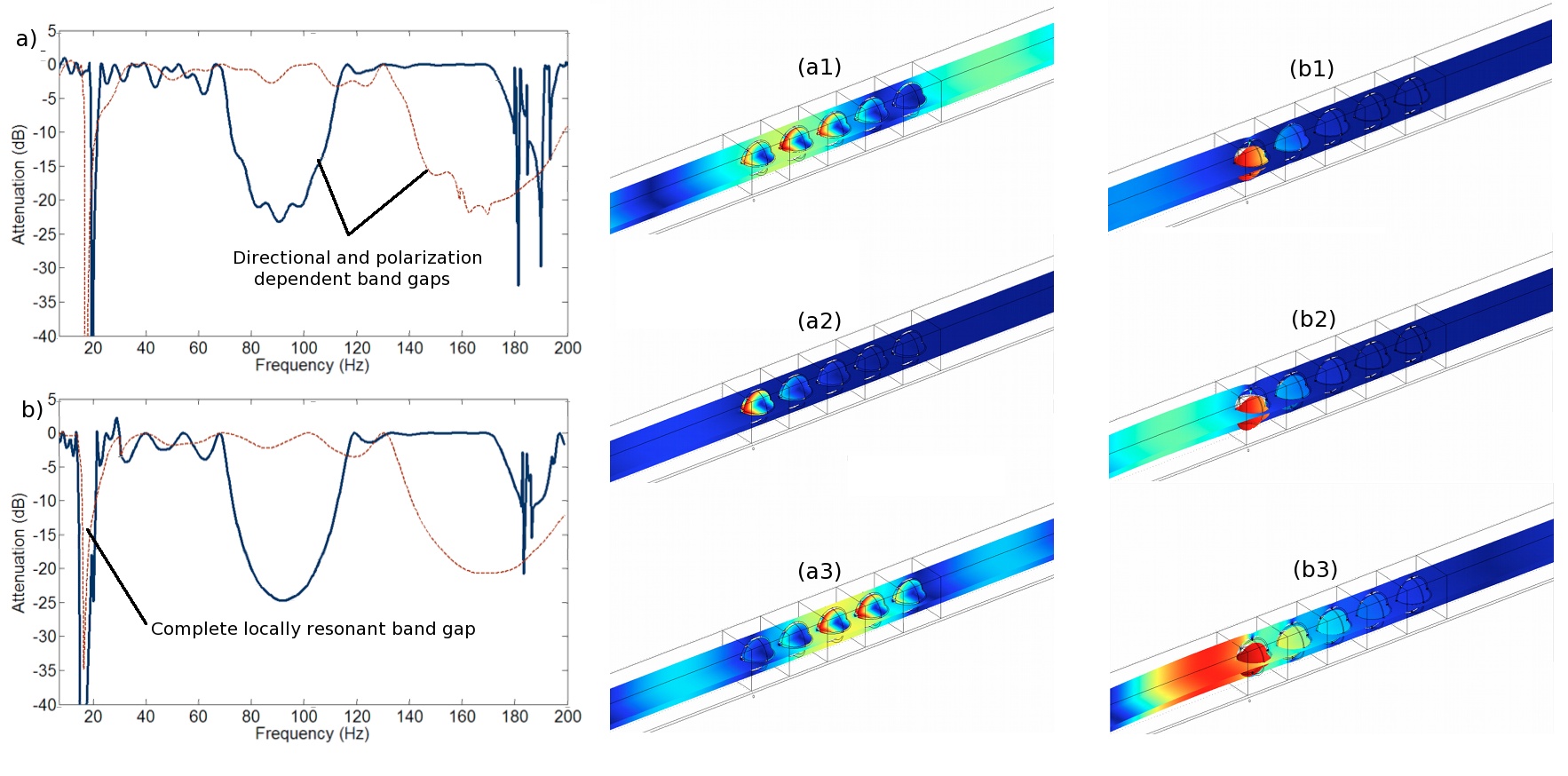}
\caption{Retreaved transmission loss through a five-period locally resonant elastic metamaterial in the case of (a) one and (b) six connected ligaments when excited with compressional (dashed brown curve) and shear (solid blue curve) waves. (a1-a3) and (b1-b3) display the total displacement field in the case of one ligament and six ligaments respectively when excited by a shear wave at $17$ Hertz, $20$ Hertz and $23$ Hertz, respectively.
}
\end{figure*}

\begin{figure}[h!]
\centering
\includegraphics[clip,angle=0,width=90mm]{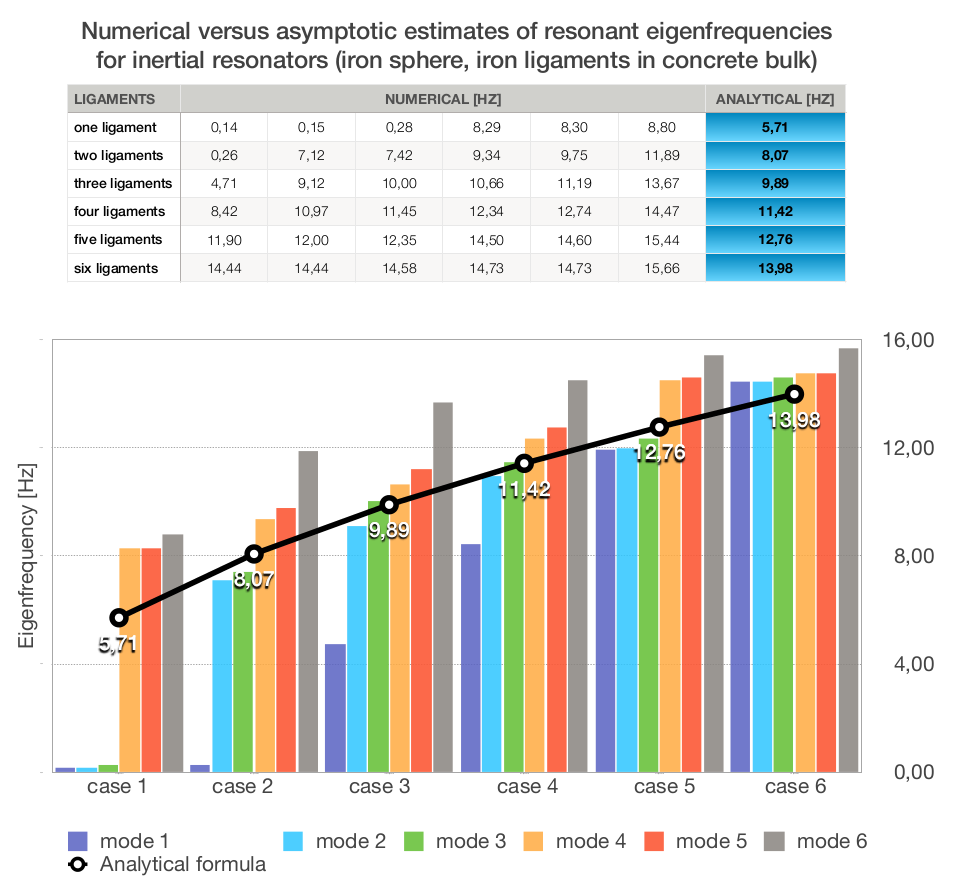}
\caption{Asymptotic (black curve) versus numerical (color bars) results for the first six local resonances associated with iron spheres of radius $R=3.7$ m attached to a bulk of concrete with one (case 1) up to six (case 6) iron ligaments of length $l_i=0.3$ m and diameter $h_i=0.2$m.
}
\end{figure}

\begin{figure*}[t]
\centering
\includegraphics[clip,angle=0,width=150mm]{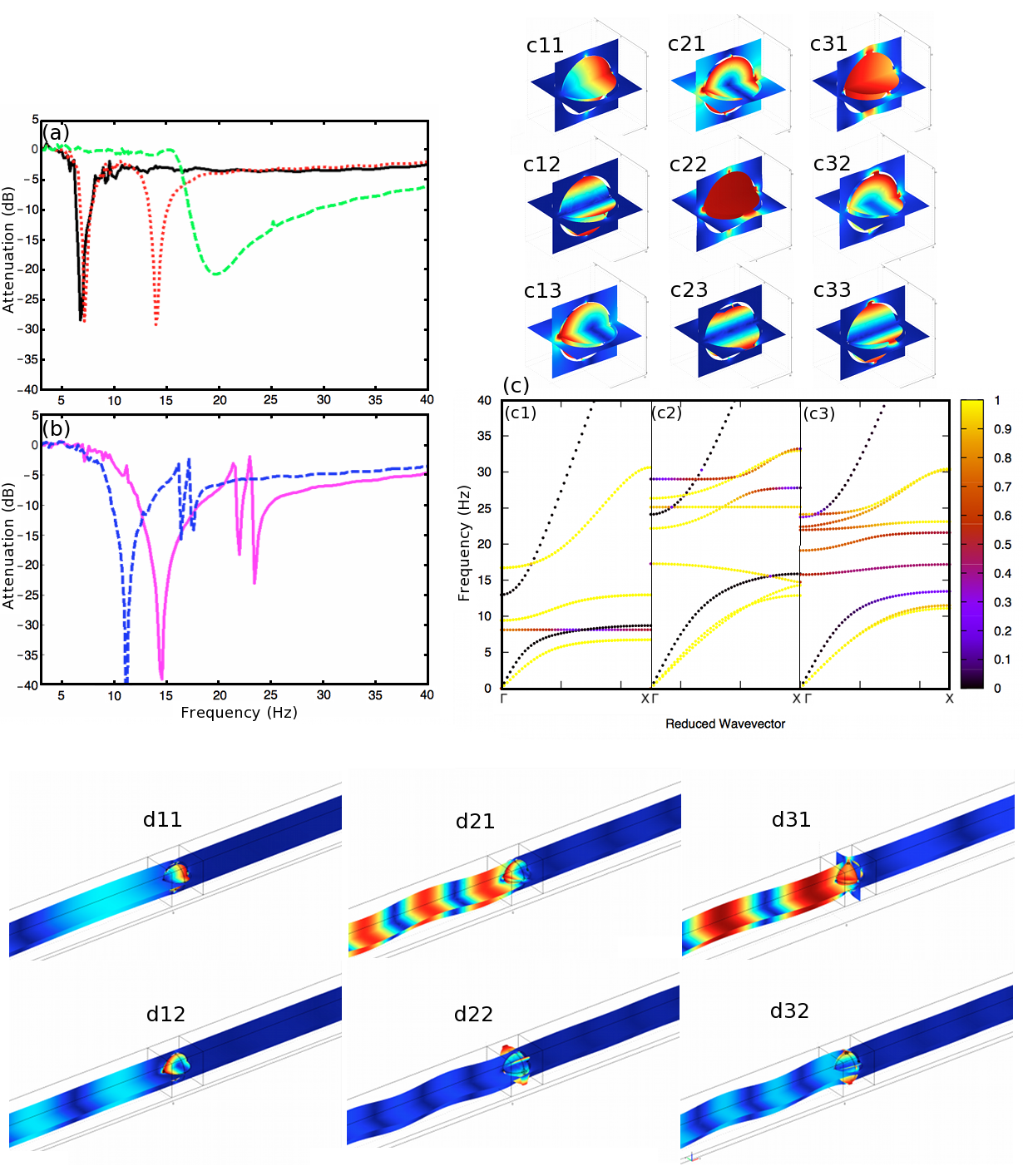}
\caption{Transmission loss response around the resonance frequency of an iron sphere of radius $R=0.37$m attached to a bulk of soil with six ligaments of polymeric material of density $1.2$ $10^3$ g.m$^{-3}$ and Young's modulus $1.1$ $10^{9}$ Pa with (a) 1/1/1/1/1/1 configuration (black line), 4/1/1/1/1/1 configuration (red line) and 1/1/4/4/8/8 configuration (green line). (b) same as (a) in the case of 1/1.5/2/2.5/3/3.5 and 1/2/3/4/5/6 configurations (blue and magenta, respectively). (c) Dispersion curves taking into account the polarization state of shear elastic waves in the case of 4/1/1/1/1/1, 1/1/4/4/8/8 and 1/2/3/4/5/6 configurations, respectively.  (c11, c12, c13, c21, c22, c23, c31, c32 and c33) and (d11, d12, d21, d22, d31, and d32) are shear bending and rotational eigenmodes that respond to a shear wave excitation at two different frequencies in the three cases, respectively. The notation $h_1$/$h_2$/$h_3$/$h_4$/$h_5$/$h_6$, $h_i$, $i=1,..,6$ represents the diameter of the i-th ligament expressed in units of $10^{-2}$ m.
}
\end{figure*}   

We start by analyzing propagation of bulk waves through a homogeneous and isotropic elastic medium, with Comsol Multiphysics, applying in this finite element software Floquet-Bloch boundary conditions on either sides of a periodic cell of side length $a=10$ m in order to compute band diagrams. In this case, it is well-known that compressional and shear waves display non dispersive curves, going to zero in a linear fashion. However, in structured media, the propagation velocity directly depends on frequency.
If we engineer some air holes (i.e. stress-free cavities without resonators placed therein) in the continuum we open Bragg band gaps, which are polarization and direction dependent band gaps.
In principle, these can be used to protect the buildings from seismic wave damages, bearing in mind they would stop mostly shear polarized waves with a frequency larger than $50$ Hertz, and a wavelength smaller than $40$ m, due to the local wave velocity. Nevertheless, the frequency range suggested by the S.D.O.F.O model is estimated to be less than $50$ Hertz (well below 5 Hertz for the first mode of the building), which makes these Bragg band gaps barely useful.

We then led numerical simulations to shed light on different vibration eigenmodes associated with an array of heavy ($7$ meters in diameter) iron spheres connected to a concrete medium via one or six tiny iron ligaments (length $0.8$ m and diameter $0.6$ m and resp. length $0.3$ m and diameter $0.2$ m) and placed inside $4$-meter radius hollow spheres. Such configurations display obviously as it has been reported in the literature \cite{achaoui1,achaoui2,achaoui3}, phononic Bragg band gaps when the wavelength and the array pitch are of the same order of magnitude and locally resonant band gaps in the subwavelength regime (figures 2-a , b, c and d). If the phononic Bragg band gap can be easily enlarged by acting on geometrical and physical properties of the inserted inclusions, the locally resonant band gaps are usually conditioned by the quality factor of the involved resonators. We consequently have to choose between the damping efficiency and the frequency bandwidth. Furthermore, it is important to mention that resonators that cross the wave pathway cannot automatically couple to the continuum \cite{achaoui4}. In figure 2, modes e1 and e3 (asymmetric bending along (Oy) axis and rotating around (Ox) axis, respectively), which are identified in the very low frequency regime ($1$ Hertz) in the band structures are examples of such phenomenon: no band gaps can be opened up accordingly. This array of resonators with only one ligament displays a tiny locally resonant stop band around $19$ Hertz thanks to the compressional mode e2 and the rotational mode e4. The substitution of the spheres of diameter $7$ m with one ligament of length $0.8$ m and diameter $0.6$ m by spheres of diameter $7.4$ m with six ligaments of length $0.3$ m and diameter $0.2$ m, makes it possible to shift the e1 and e3 modes (figure2-e) towards higher frequencies.  This gives rise to modes f1 and f3 (figure2-f). This configuration enables a local enlargement of the band gap as shown in figures 2-c, 2-d.

The transmission loss over a finite structure made of five rows and the displacement field maps of the aforementioned configurations under a shear excitation are presented in figure 3. Periodic boundary conditions are applied in the two directions perpendicular to the wave propagation to prevent beam's and plate's waves. We have numerically checked that the six-ligament-resonator array leads to wider locally resonant bandgaps in comparison with the directional and polarization dependent band gaps induced by arrays of hollow spheres. A particular attention is brought to the displacement field maps (figures 3-a1, a2, a3, b1, b2 and b3). Since we have a shear excitation, the compressional resonant modes are not excited in both cases (one and six-ligament-resonator periodic structures). Consequently, the rotational mode is responsible for the wave attenuation in the case of one-ligament-resonators and in theory (according to the band structures) both the rotational mode and the bending shear mode should contribute to the attenuation. Nevertheless, at different frequencies inside the gap, only the bending shear wave appears to be excited. The rotational mode seems to be totally overwritten by the latter. This information is of foremost importance since it reveals that all the mechanisms are not fully exploited and an optimization can be performed so as to achieve a wider range of frequencies for which wave propagation is prohibited.

\begin{figure*}[t]
\centering
\includegraphics[clip,angle=0,width=140mm]{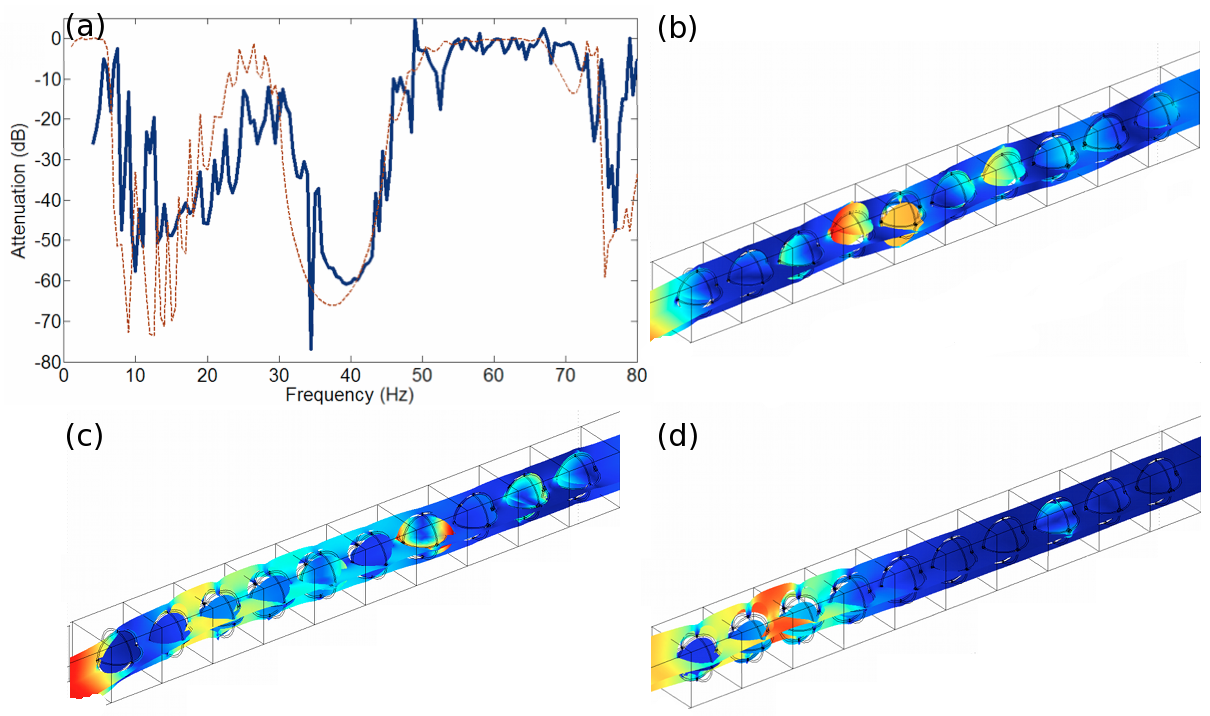}
\caption{a) Transmission loss through a ten array periodic structure made of inertial resonators using chirped ligaments. b), c) and d) are examples of excitations at three different frequencies. In b) pure and hybrid bending modes are responsible for damping. In c) pure bending modes along (Oz) axis, pure rotational modes around (Oz) axis and pure rotational modes around (Oy) axis, are responsible for damping. In d) pure bending and rotational modes around (Ox) axis contribute to the attenuation of the wave propagation.
}
\end{figure*}

In order to optimize the location of low frequency resonances associated with the stop band, we use an asymptotic estimate of eigenfrequencies
associated with rotational modes of the heavy spheres (adapted from \cite{mazya,guenneau1}):
\begin{equation}
f=\frac{1}{4\pi}\sqrt{\frac{\mu}{4\rho}\frac{\lambda+\mu}{\lambda+2\mu}\sum_{i=1}^N{\frac{h_i^2}{R^3 l_i}}}
\end{equation}
where $R$ is the radius of a sphere, $h_i$ and $l_i$ are the diameter and length of the thin ligaments
attached to it. Also, $\rho$ is the density, $\lambda$ and $\mu$ are the Lam\'e coefficients of the ligaments.

One can see that the aspect ratio $h_i^2/l_i$ scales down the resonant frequency when the ligaments become thinner and thinner, what allows us to neatly tune down the frequency of
the stop band by playing with thin ligaments' size. Moreover, the formula tells us that adding more ligaments results in shifting up the resonant frequency. One can therefore fine tune the location of stop bands
simply by optimizing aspect ratios and number of ligaments. Ligaments need not have the same size in the formula, and it turns out that considering ligaments of
different sizes breaks the degeneracy of eigenfrequencies, what leads to multiple closely located resonances enlarging the width of the range of operating damping frequencies as we will explain in detail in the next section. 

In order to evaluate the accuracy of the formula, we carried out a comparative study between the numerical and analytical results of resonant eigenfrequencies (figure 4) for the case where the resonator ($10\times 10\times 10$ m$^3$) components are from iron (the ball and the ligaments) and concrete (the bulk). We can clearly observe that the asymptotic results (black curve) approximate quite well the eigenmodes responsible for opening low frequency total band gaps regardless of the number of ligaments. For further details please refer to Supplemental materials, where we test the asymptotic formula against finite element computations not only for a concrete host matrix but also for a soft one such as soil and various types of ligaments (rubber and iron), in which case one needs to introduce an effective length for the ligaments.

\section{Achieving broadband seismic wave damping}

Bearing in mind that earthquake shielding is a potential application of this theoretical work, we decided to adapt our structure's physical and geometrical parameters to those used in \cite{krodel} where numerical predictions have been validated experimentally at a reduced scale. The elementary cell is now $1$ meter in sidelength and the physical parameters for the ligaments are for a polymeric material (Vero Blue) with density $\rho=1.2\; 10^{3}$ kg.m$^{-3}$ and Young's modulus $E=1.1 \; 10^9$ Pa. This material
could be replaced by commercially available construction rubber bearings for civil engineering applications. The host medium is made of soil with density $\rho=1.3\; 10^{3}$ kg.m$^{-3}$ and Young's modulus $E=20 \; 10^6$ Pa.
We first evaluated the transmission signature of one resonator for different configurations. We would like to stress here that only the ligaments' width are changed, all the other parameters remain unchanged. A shear elastic excitation applied to the case of a resonator with six identical ligaments $1 \; 10^{-2}$ m in diameter as tested in the previous section give rise to a wave attenuation of  $28$ dB around $6.5$ Hertz (black continuous line in figure 5). The attenuation is due to the bending shear mode as we explained previously. When we increase the diameter of one ligament up to $4 \; 10^{-2}$ m for example (the other ligaments being kept unchanged), another dip shows up around $14.5$ Hertz and the first one remains almost unchanged (red dotted lines). Figures 5-(d11) and (d12) display the displacement field maps at $6.5$ Hertz and $14.5$ Hertz, respectively. We can identify the bending shear mode ((d11), asymmetric because of a large ligament) and the rotational mode (d12). At this stage we succeeded in exploiting both the bending shear mode and one rotational mode. Furthermore, the position of this rotational mode can be easily tuned by changing the width of one ligament. When we take a look at the dispersion curves (colored to highlight the shear modes) (figure 5-c1), we note three colored flat modes at $6.5$ (c11), $8$ (c12) and $14.5$ Hertz (c13). We note the correspondence between the eigenmodes (c11) vs (c13) and (d11) vs (d12). However, the mode (c12) is totally overwritten by (c11) in the transmission spectrum. We continue our analysis by changing the diameter of four ligaments in such a way that the two of them oriented towards (Ox) remain with a diameter of $1 \; 10^{-2}$ m while the two oriented towards (Oy) are set at $4 \; 10^{-2}$ m and the other two are set at $8 \; 10^{-2}$ m. The transmission curve (green dashed line) represents a relatively large dip around $20$ Hertz which is confirmed by the polarization state band structures (figure 5-c2). It is worth noting that the presence of resonant modes around the band gap allows for an attenuation over several Hertz. We notice that the phenomenon is ruled not only by the pure rotational mode (see figures 5 (c23), (d22)) but also by a hybrid mode (see figure 5 (d21)) that results from a combination between modes in figures 5 (c21) and (c22). Last but not least, we demonstrate that if now the unit cell is totally asymmetric, which can be achieved by applying an incrementation step of one $1 \; 10^{-2}$ m from one ligament to another, the resonant modes are further separated (figure 5-(b), magenta continuous line, and (c3)). Consequently, the bending shear mode and the two purely rotational modes can be exploited. This configuration can also be tuned by applying different incremental steps (blue dashed line of figure 5-(b) with a step of $0.5$ Hertz).

These multiple-resonance units with a judicious choice of physical and geometrical parameters could be considered as an efficient solution to prohibit elastodynamic wave propagation in potential applications such as noise and vibration reduction or in earthquake protection. Indeed, unlike devices involving conventional resonators, wherein the resonance frequency is linked to the bulk with a very tiny band gap feature, a periodic structure with ten rows consisting of miscellaneous resonators with almost the same bulk
can exhibit an ultrawide complete stop band as shown in figure 6-(a). At three different frequencies (figure 5-(b), (c), (d)) one can identify the units contributing to the prohibition of the wave propagation via their localized eigenmodes.

\section{Conclusion}
To conclude, we would like to stress that potential applications of our study are in seismic shields against surface elastodynamic waves with short wavelengths that occur in sedimentary basin. Our proposal of an earthquake shield with a periodic array of pitch $1$ meter consisting of iron spheres of diameter $0.74$ meter connected via thin rubber ligaments to a bulk of concrete inserted underneath or around the foundations of large human infrastructures, which reflects all elastodynamic wave polarizations falling within a large stop band that occurs between $8$ and $49$ Hertz depending upon the ligaments which we used, should therefore find some civil engineering applications in seismic regions with soft sedimentary soil. Our proposal is reminiscent of recent works on acoustic resonators for seismic wave reflection and damping to be inserted around, undernearth or within \cite{kim,shi,finocchio,krodel}, buildings. However, we note that in our case, low frequency stop bands rely upon inertial resonators \cite{guenneau2,bertoldi,achaoui4}, which were first conceptualized by Profs. Davide Bigoni and Alexander Movchan. Let us point out that in \cite{brule2}, wave shielding is induced by Bragg stop bands with boreholes in sedimentary soil, hence this type of seismic metamaterial necessarily shares some of the features of phononic crystals. A first experimental step towards seismic metamaterials with subwavelength stop bands has been achieved with forest of trees in \cite{andrea1}, but for surface Rayleigh waves at relatively high and narrow band frequencies. In contradistinction, in the current proposal, ultra-low frequency stop bands induced by a locally resonant structure buried within the soil achieves some bulk elastic wave damping for all polarizations. We hope that our  study will foster experimental efforts in meter and decameter scale metamaterials for surface and volume mechanical wave control. In the course of our paper's revision we found some related analysis of low frequency broadband stop band metamaterials in \cite{daraio}. 

Y. Achaoui and S. Guenneau acknowledge funding of European Union (ERC grant ANAMORPHISM).
*Y. Achaoui and B. Ungureanu are equal contributors.


\begin{thebibliography}{99}
\bibitem{Reitherman}
Reitherman R., Earthquakes and Engineers: An International History. Reston, VA: ASCE Press, 2012.
\bibitem{symans}
Symans M.D., Charney, F.A., Whittaker A.S., Constantinou M.C., Kircher C.A., Johnson M.W. and R.J. McNamara R.J., Energy dissipation system for seismic application: current practice and recent developments, Journal of Structural Engineering 10.1061/(ASCE) 0733-9445, 134:1(3) (2008).
\bibitem{brule1}
Br\^ul\'e S. and Javelaud E., in Proceedings of the 9th Annual International Conference on Urban Earthquake Engineering (Tokyo Tech CUEE, Tokyo, 2012), p. 497 (2012).
\bibitem{brule1b}
Br\^ul\'e S., Javelaud E., Guenneau S., Enoch S. and Komatitsch D., in
Proceedings of the 9th International Conference of the Association for Electrical Transport and Optical Properties of Inhomogeneous Media
ETOPIM 9, 2-7 september2012,  Marseille (2012).
\bibitem{brule1a}
Br\^ul\'e S., Javelaud E. and Marchand M., Chimneys health monitoring during a nearby heavy dynamic compaction site, in Journ\'ees Nationales de G\'eotechnique et de
G\'eologie de l'Ing\'enieur JNGG2012, 4-6 july 2012, Bordeaux, France, Tome II, 919-926 (2012).
\bibitem{gmur}
Gmur T., Dynamique des structures. Analyse modale numerique, Presses Polytechniques et Universitaires Romandes, Lausanne, Switzerland, 2008.
\bibitem{chopra}
Chopra A., Dynamics of structures. Theory and applications to earthquake engineering, Pearson Prentice-Hall, Upper Saddle River, NJ, 2012.
\bibitem{brule2}
Br\^ul\'e S., Javelaud E., Enoch S. and Guenneau S.,
Phys. Rev. Lett. 112, 133901 (2014).
\bibitem{andrea1}
Colombi, A., Craster, R.V., Gueguen, P., Guenneau, S. and Roux, P.,
Scientific Reports (in press) 
\bibitem{brule3}
Sheng P.,
Physics 7, 34 (2014).
\bibitem{sheng}
Liu Z., Zhang X., Mao Y., Zhu Y. Y., Yang Z., Chan C. T. and Sheng P., Science 289, 1734 (2000).
\bibitem{djafari}
Goffaux, C., Sanchez-Dehesa, J., Levy Yeyati, A., Lambin, Ph., Khelif, A., Vasseur, J.O. and Djafari-Rouhani B.,
Phys. Rev. Lett. 88, 225502 (2002).
\bibitem{mechanical1}
Martinsson P.G. and Movchan, A.B.,
Quat. J. Mech. Appl. Math. 56, 45-64 (2003).
\bibitem{mechanical2}
Goffaux, C. and Sanchez-Dehesa, J.,
Physical Review B, 144301 (2003).
\bibitem{li}
Li, J. and Chan, C.T.,  Phys. Rev. E, 70, 055602 (2004).
\bibitem{movchan}
Movchan, A.B. and Guenneau, S., Phys. Rev. B, 70, 125116 (2004).
\bibitem{fang}
Fang, N., Xi, D. J., Xu, J. Y.,Ambrati, M., Sprituravanich, W., Sun, C. and Zhang, X., Nat. Mater. 5, 452 (2006).
\bibitem{sar}
Guenneau, S., Movchan, A.B., Petursson, G. and Ramakrishna, S.A.,
New Journal of physics 9 (11), 399 (2007).
\bibitem{milton}
Milton, G.W. and Willis, J.R.,
Proc. Lond. Roy. Soc. 463, 855-880 (2007).
\bibitem{theocharis}
Theocharis, G., Kavousanakis, M., Kevrekidis, P.G., Daraio, C., Porter, M.A. and Kevrekidis, I.G.,
Phys. Rev. E 80, 066601 (2009).
\bibitem{hsu}
Hsu J.C.,
Journal of Physics D: Applied Physics 44 (5), 055401 (2011). 
\bibitem{craster}
Craster, R.V. and Guenneau, S., Acoustic Metamaterials, Springer Verlag, 2012.
\bibitem{mechanical3}
Martin,  A., Kadic, M., Schittny, R., Buckmann, T. and Wegener, M., Physical Review B 86, 155116 (2012).
\bibitem{fleury}
Fleury, R., Sounas, D.L., Sieck, C.F., Haberman, M.R. and Alu, A.,
Science 343, 516 (2014). 
\bibitem{achaoui1}
Benchabane, S., Gaiffe, O., Ulliac, G., Salut, R., Achaoui, Y. and Laude, V.,
Appl. Phys. Lett. 98(17), 171908 (2011).
\bibitem{achaoui2}
Khelif, A., Achaoui, Y., Benchabane, S., Laude, V. and Aoubiza B.,
Physical Review B 81 (21), 214303 (2010).
\bibitem{achaoui3}
Achaoui, Y., Khelif, A., Benchabane, S.,  Robert, L. and Laude, V.,
Physical Review B 83 (10), 104201 (2011).
\bibitem{mazya}
Kozlov, V.A., Maz'ya, V.G. and Movchan A.B.,
Asymptotic analysis of fields in multi-structures, Oxford Research Monographs, Oxford University Press 1999
\bibitem{guenneau1}
Guenneau S., Movchan A. B. and Movchan N. V., Physica B 394, 141-144 (2007).
\bibitem{guenneau2}
Bigoni, D., Guenneau, S., Movchan, A.B. and Brun, M., Phys. Rev. B 87(17), 174303 (2013).
\bibitem{bertoldi}
Wang, P., Casadei, F., Shan, S., Weaver, J.C.  and Bertoldi, K., Phys. Rev. Lett. 113, 014301 (2014).
\bibitem{andrea2}
Colombi, A., Roux, P. and Rupin, M., 
The Journal of the Acoustical Society of America 136 (2), EL192-EL198 (2014).
\bibitem{achaoui4}
Achaoui, Y., Diatta A., and Guenneau, S., Appl. Phys. Lett. 106, 223502 (2015).
\bibitem{jensen}
Sigmund, O. and Jensen J. S.,
Philosophical Transactions of the Royal Society of London. Series A
361(1806), 1001-1019 (2003)
\bibitem{xiao}
Xiao, Y., Wen, J. and Wen, W., New J. Phys. 14, 033042 (2012). 
\bibitem{kim}
Kim, S. H. and Das, M. P., Mod. Phys. Lett. B 26, 1250105, (2012).
\bibitem{shi}
Shi, Z., Cheng, Z. and Xiang, H., Soil Dyn. Earthquake Eng. 57, 143-151 (2014).
\bibitem{finocchio}
Finocchio, G., Casablanca, O., Ricciardi, G., Alibrandi, U., Garesci, F., Chiappini, M. and Azzerboni, B.,
App. Phys. Lett. 104, 191903 (2014).
\bibitem{krodel}
Krodel, S., Thome, N. and Daraio, C.
Ex. Mech. Letters 4, 111-117 (2015).
\bibitem{daraio}
 Matlack, K.H., Bauhofer, A., Krodel, S., Palermo, A. and Daraio, C.
arxiv:1511.09465v1
\end{thebibliography}
\end{document}